# Measurement and Interpolation for Data-Driven Pressure Distribution Rendering on a Finger Pad

Kazuya Sase[1], Rei Onodera[2], Hikaru Nagano[3], and Masashi Konyo[4]

[1] Faculty of Engineering, Tohoku Gakuin University, Sendai, Japan
[2] Graduate School of Engineering, Tohoku Gakuin University, Sendai, Japan
[2] Graduate School of Science and Technology, Kyoto Institute of Technology, Kyoto, Japan
[2] Graduate School of Information Sciences, Tohoku University, Sendai, Japan
(Email: sase@mail.tohoku-gakuin.ac.jp)

**Abstract ---** We propose a data-driven pressure distribution rendering method that uses the interpolation of experimentally obtained pressure values. The pressure data were collected using a pressure sensor array and prediction was performed using linear interpolation assuming a pressure distribution dependent on the pushing displacement and contact angle. In our demo, a Leap Motion Controller was used to capture user input, and the predicted distribution, based on this input, is displayed on a liquid crystal screen in real time. The proposed prediction model was found to be fast and reproduced the measured data well.

**Keywords: pressure distribution, haptic rendering, data-driven method**

## 1 Introduction

Humans perceive the sense of touch through contact with objects on the skin surface. Mechanoreceptors are located beneath the skin, and several types of receptors encode mechanical phenomena as tactile signals. They are densely distributed on the skin, and the receptor density is particularly high on the finger pads. The dense distribution of receptors allows observing the mechanical phenomena occurring on the surface of the skin.

Reproducing the spatial distribution of haptic stimuli in virtual reality is crucial for increasing the reality of the created experience and convincing people of the existence of virtual objects. The artificial creation of spatially distributed haptic stimuli has been studied extensively. For example, researchers have developed a pin-array-type haptic display that can display spatially distributed vibrations [1] or the shape of an object's surface [2]. Moreover, haptic distribution displays based on air pressure [3][4], electrotactile stimuli [5], and airborne ultrasound focusing [6] have also been developed.

Haptic rendering, which is the software's ability to generate signals for haptic displays, is also important. Hirota et al. proposed using the finite element method (FEM) to simulate the entire hand [7] and generated a pressure distribution when a finger contacted a rigid object [4]. We also used FEM to compute the deformation of the soft body with the finger modeled as a rigid body. Although FEM-based approaches can simulate a plausible pressure distribution, the accuracy of the obtained results has not yet been verified. Furthermore, it is unclear how a finger structure can be accurately created using tissue layers, nails, and bones. Moreover, computing complex structures using FEM in real time is challenging.

Machine learning techniques have also been used to achieve cost-effective haptic rendering. Culbertson et al. proposed a data-driven method for reproducing vibrations caused by stroking different textured surfaces using a pen [8]. Hover et al. developed a force-rendering method based on impedance measurements and the interpolation of viscoelastic materials [9]. However, no data-driven method has been reported for reproducing the pressure distribution caused by contact between a finger and an object.

Here, we propose a data-driven pressure distribution rendering method that relies on data collected from experiments with real fingers. We used a pressure array sensor for data collection and interpolated the measured values using the finger displacement and contact angle. This paper describes the results of this study.

## 2 METHODS

### 2.1 Overview

Our method includes data collection to determine the source of pressure distribution prediction. Data collection is required in situations where the finger dimensions, material properties, and object shapes vary. In this study, we focused on the situation in which a person's index finger touches the surface of a rigid flat plane. Additionally, considering that the pressure distribution depends on the displacement and angle of the finger, the goal was set to create a predictive model as shown in Fig. 1.

### 2.2 Data Collection

A capacitance-type pressure sensor array comprising 1.5-mm$^2$ sensor elements (TACTARRAY MODEL #8058, Pressure Profile Systems, Inc.) was used to measure the pressure distribution. The sensors are aligned in 16 rows and 15 columns. The full-scale pressure was 82.87 kPa. The measurement principle is based on the response of the capacitance to a change in the distance between the two electrodes, and it is well known that this method is effective in capturing small pressure changes.

We conducted an experiment to measure the pressure distributions under conditions of finger displacement in the pushing direction and a change in the contact angle. This experiment involved a participant (35-year-old male, one of the authors). The experimental equipment used for data collection is shown in Fig. 2. The participant fixed the nail side of their index finger onto three rigid plates prepared and angled at 15°, 30°, and 45° with respect to the horizontal plane. The finger was displaced using a precision z-stage. The displacement baseline was set as the point at which the measured pressure exceeded the threshold of 0.5 kPa. After establishing the baseline, the finger was displaced in 0.5 mm increments until it reached 1.5 mm. We stopped at each measurement point to obtain the static values. Subsequently, the pressure values measured at each measuring point were averaged to obtain a representative displacement value.

### 2.3 Interpolation

As a preliminary attempt, we used simple bilinear interpolation to predict the pressure values between the measured points. The measured pressure distribution was 16×15, which was correlated with the displacement and contact angle. Each pressure value in the distribution was

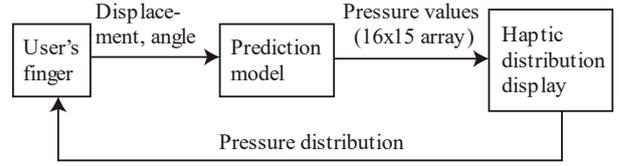

Fig. 1 Overview of haptic rendering

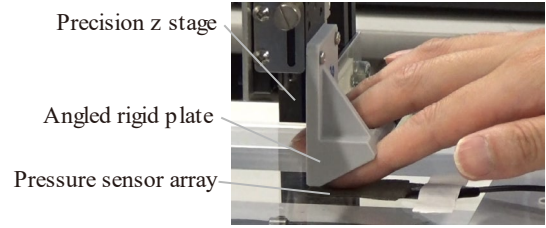

Fig. 2 Experimental setup

interpolated using the following equation:

$$p_i(z, \theta) = p_i(z_0, \theta) + \frac{p_i(z_1, \theta) - p_i(z_0, \theta)}{z_1 - z_0}(z - z_0),$$

with

$$p_i(z_0, \theta) = p_i(z_0, \theta_0) + \frac{p_i(z_0, \theta_1) - p_i(z_0, \theta_0)}{\theta_1 - \theta_0}(\theta - \theta_0),$$

$$p_i(z_1, \theta) = p_i(z_1, \theta_0) + \frac{p_i(z_1, \theta_1) - p_i(z_1, \theta_0)}{\theta_1 - \theta_0}(\theta - \theta_0),$$

where $p_i$ denotes the interpolated pressure value of $i$-th sensor element; $z$ and $\theta$ denotes the input displacement and contact angle, respectively; and $z_0$, $z_1$, $\theta_0$, and $\theta_1$ denote the displacements and contact angles at the measurement points ( $z_0 < z_1$ and $\theta_0 < \theta_1$ ). $p_i(z_0, \theta_0)$, $p_i(z_0, \theta_1)$, $p_i(z_1, \theta_0)$ and $p_i(z_1, \theta_1)$ are measured values and thus known values. By the above equation, the values of $p_i(z, \theta)$ in the range between $z_0$ and $z_1$, or between $\theta_0$ and $\theta_1$, can be interpolated. This interpolation was applied to all pressure elements independently.

## 3 RESULTS AND DISCUSSION

### 3.1 Measured data and its interpolation

The measured pressure distributions of the participant are shown in Fig. 3, and interpolation was performed based on these data. Fig. 4 shows three examples of the interpolated pressure distributions, where the values of the input variables were selected between the sample points.

### 3.2 Real-time pressure distribution calculation

To confirm the adequacy of the approach, we performed real-time pressure distribution interpolation with user displacement input using a hand-tracking device, the Leap Motion Controller (Leap Motion Inc.).

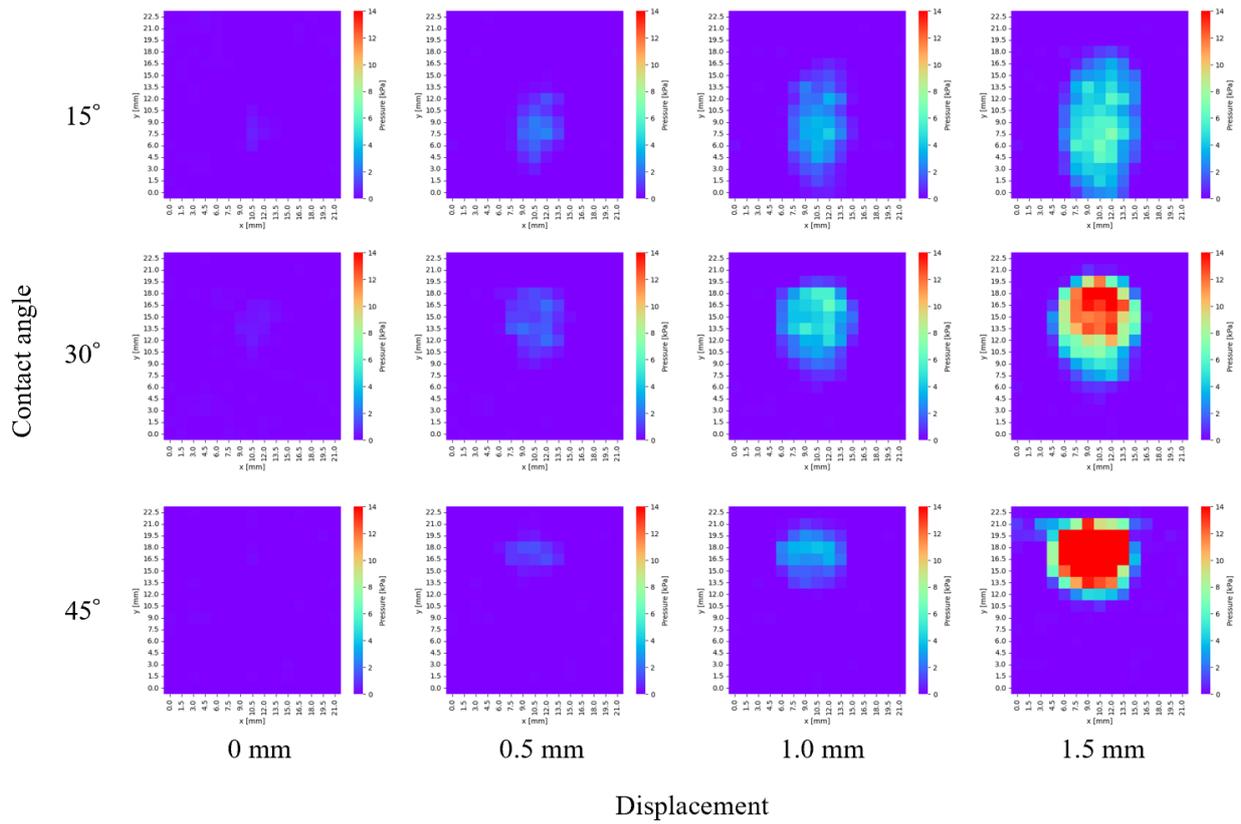

Fig. 3 Measured pressure distributions

An overview of the system is shown in Fig. 5. The finger displacement was captured using the device, and the predicted pressure distribution was visually rendered on a liquid crystal display (but not haptically displayed at this time). The predicted pressure distributions were interpolated continuously from the measured values. The prediction was calculated immediately (less than 10 μs) using a computer.

### 3.3 Discussion

As expected, the prediction successfully interpolated the measurement values, and the computation time was minimal. In this experiment, only two inputs were considered. Therefore, we could not account for all finger movements in the pressure prediction. This can be improved by increasing the number of input variables, preferably by three translations and three rotations.

This simple interpolation method completely fits the measured values, demonstrating that various measurement errors, such as outliers, can have a significant impact on the prediction. Therefore, other machine learning techniques should be used to address this overfitting issue.

### 4 Conclusion

In this study, we developed a pressure distribution prediction method based on linear interpolation of experimentally measured values. This method is simple and fast, and reproduces the measured data well. The method was verified using only two input variables; however, it suffers from overfitting. Therefore, the fitting method needs to be improved to reproduce complex haptic scenarios. Furthermore, data collection can be performed using a less-constrained experimental setup. These issues should be addressed in future studies.

### Acknowledgement

This work was supported by JSPS KAKENHI, Grant Numbers JP22K17936 and JP21H04542.

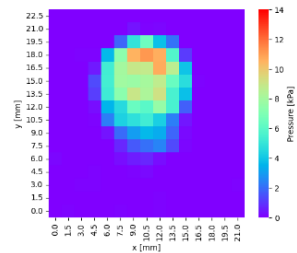

(a) $z = 1.25$ [mm], $\theta = 30$ [°]

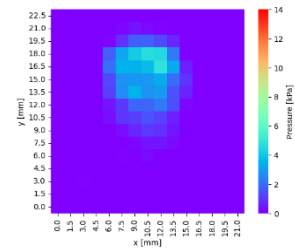

(b) $z = 1.0$ [mm], $\theta = 37.5$ [°]

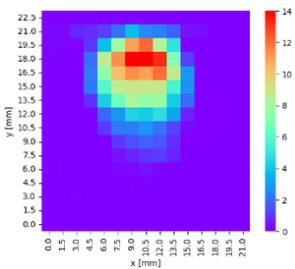

(c) $z = 1.25$ [mm], $\theta = 37.5$ [°]

Fig. 4 Interpolated pressure distribution

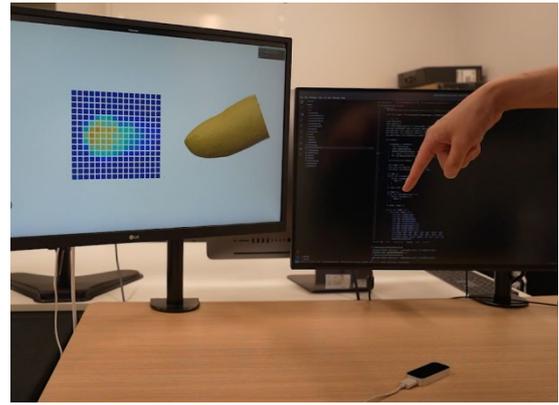

Fig. 5 Real-time pressure distribution calculation